\documentclass[aps,pre,twocolumn,amsfonts,amssymb,superscriptaddress,preprintnumbers,showpacs]{revtex4}
\usepackage{placeins}
\usepackage{graphicx}
\usepackage{amsmath}
\usepackage{times}
\usepackage{colordvi}
\usepackage{epsfig}

\newcommand{\IGN}[1]{}
\newcommand{\ZWR}[1]{}

\begin{document}
\title{Cyclic dominance and biodiversity in well-mixed populations}
\author{Jens Christian Claussen}  \affiliation{Institut f{\"u}r Theoretische Physik und Astrophysik, Christian-Albrechts Universit{\"a}t, Olshausenstra{\ss}e 40, 24098 Kiel, Germany}
\author{Arne Traulsen} 
\affiliation{Max-Planck-Institute f{\"u}r  Evolutionsbiologie,  August-Thienemann-Str. 2, 24306 Pl{\"o}n, Germany} 
\date{June 19, 2007; revised
 October 26, 2007
}

\preprint{Physical Review Letters, in print (2008)}

\begin{abstract}
Coevolutionary dynamics
is investigated in chemical catalysis, biological evolution,
social and economic systems.
The dynamics of these systems can be 
analyzed within the unifying framework of evolutionary game theory.
In this Letter, we show that even in well-mixed finite populations, 
where the dynamics is inherently stochastic,
biodiversity is possible
with three cyclic dominant strategies.
We show how the interplay of evolutionary dynamics, 
discreteness of the population, and the nature of the interactions
influences the coexistence of strategies. 
We calculate a critical population size above which coexistence is likely.
\end{abstract}
\pacs{
87.23.-n, 		
89.65.-s 		  
02.50.Ey 		
}
\maketitle

Coevolution with cyclic dominance can 
lead to oscillatory, chaotic and stochastic dynamics. 
For example, such cycles can be found 
in biology 
\cite{magurran,sinervo96,Zam00,kerr,kirkup,dawkins,bastolla05}
or 
in social and economic systems 
\cite{helbing,loners}.
The simplest example for such a cyclic dominance 
is the well known children's game rock-paper-scissors,
where rock crushes scissors, scissors cuts paper and paper wraps rock. 
Different biological realizations
of this system have been observed. 
For example, a cyclic dominance of
three male strategies has been reported in lizards
\cite{sinervo96,Zam00}:
Orange-throated males establish 
large territories holding several females. These populations are invaded by males with 
yellow-striped throats, which do not contribute to the defense of the territory
but sneak on the females.  
Such a population of yellow-striped males can be invaded by blue-throated males, which defend territories large enough to hold one female which they defend against sneakers. 
Once yellow-striped sneakers are rare, it is advantageous to defend a large territory with several females and the cycle starts with orange-throated males again. 
Another example 
is the competition between different strains of E.coli.
Kerr {\sl et al.} 
\cite{kerr,kirkup} 
observed that 
cyclic dominance
leads
to biodiversity in spatial systems,
whereas two strategies go
extinct in mixed systems after a short time. 
Such systems have been analyzed 
by evolutionary game theory
in great mathematical detail 
\cite{hofbauer98,szaboreview,nowakbook,szaboczaran01%
,czaran02%
,reich07a%
,reich07b%
}. 
Based on the replicator dynamics
describing the dynamics in an infinite population, general conclusions on the nature of the payoff matrix can be made from the observation of fast extinction. 
How does this picture change if we tackle the 
more realistic case
of stochastic dynamics
in a finite population
\cite{moran,nowak04,taylor04}?
Depending on population size and the underlying microscopic process
\cite{TCH05,CT05,TCH06,reichenbach06}, the resulting dynamics can be very different
from the the replicator equation results. 

{\sl Definition of the model.~--}
We first concentrate on the simplest example
of cyclic rock-paper-scissors dynamics,
in which all three strategies are equivalent
 \cite{sandholmEGT}. 
Thus, the game is defined by three payoffs:
(i) the payoff against a dominated strategy (set to $1$).
(ii) the payoff against a dominant strategy $-s$, which we assume to be negative. 
(iii) the payoff for a tie (set to $0$).
Thus, we obtain the payoff matrix
\begin{equation}
\label{Mmatrix}
	\bordermatrix{
		  & R & P & S \cr
		R & 0 & -s & 1 \cr
		P & 1 & 0 & -s \cr
		S & -s & 1 & 0 \cr
		}.
\end{equation}
Only for the standard choice $s=1$, we have a zero-sum game. 
An intuitive understanding how $s$ influences the game can
be obtained from discussing two relevant cases. 
Large values of $s$ will make it successful to 
avoid losing, best done with 
staying with the majority. In this case, a mixed equilibrium is unstable
and ultimately, only one strategy will survive.
For $s \approx 0$, it is more important to win occasionally, 
such that the mixed equilibrium can become stable.

Let us first recall the evolutionary dynamics in an infinite population described by the
replicator dynamics \cite{hofbauer98}. In the replicator equation, the frequency 
(abundance) $x_k$ of strategy $k$ changes proportional to its payoff
$\pi_k$,
\begin{equation}
\dot x_k= x_k \left( \pi_k - \langle \pi \rangle \right).
\label{rd}
\end{equation}
Here $\langle \pi \rangle $ is the average 
payoff
in the population.
We denote the frequencies of R, P, S as $x,y,z$, respectively, 
with $x+y+z=1$.
With the standard assumption that payoffs are determined from interactions with a representative subset of the population, 
we find
$\pi_R = z-s y$,
$\pi_P = x-s z$, and
$\pi_S = y-s x $.  
The average payoff is given by $\langle \pi \rangle =(1-s)(xy+xz+yz)$
vanishing for the zero-sum game with $s=1$.
Eq.~(\ref{rd}) has an interior fixed point ${\boldsymbol p}=(\frac{1}{3},\frac{1}{3},\frac{1}{3})$. 
 The determinant of the payoff matrix $d=1-s^3$ determines the dynamics of the system \cite{hofbauer98}.
For $d>0$, which is the case for $s<1$, 
\begin{equation}
H=-x y z = - x y (1-x-y)
\label{constantofmotion}
\end{equation}
 is a Lyapunov function with $\dot H <0$ and 
the interior fixed point ${\boldsymbol p}$ is
asymptotically
stable. 
For $d<0$ ($s>1$), 
${\boldsymbol p}$ is unstable and the attractor of the system
approaches a heteroclinic cycle at the boundary of the simplex $S_3$.
Finally, for the zero-sum game with $s=1$ and $d=0$ 
the function
(\ref{constantofmotion})
is a constant of motion, and the system infinitely oscillates
around ${\boldsymbol p}$. 
This is not a purely theoretical exercise: Experiments indicate $d>0$
for the Lizard system
 \cite{Zam00} and $d<0$ for the E.coli system \cite{kerr}.  
In theoretical approaches, often the restriction to $s=1$ is made, 
 although this is a special non-generic case \cite{reich07a,reich07b}. 
 It has been argued that in this case, 
limited
 mobility in a spatially extended system can promote
 biodiversity. However, for $s<1$ biodiversity is likely even in a well-mixed population if the
 population is larger than $N_c$. 
In this Letter, we give an analytical estimate for this critical population size,
 depending on $s$ and the update mechanism. 
For $s=1$ the average time to extinction scales linearly
with the population size $N$ \cite{reichenbach06}. 
For $s>1$, one expects faster extinction, as the stochastic drift and the
instability of the interior fixed point act jointly. 
In the most interesting case of $s<1$, both forces work in opposite directions:
For $N \to \infty$, the fixed point  ${\boldsymbol p}$ is stable, but a finite $N$ leads
to a drift towards the simplex boundaries. 
 
{\sl Evolutionary processes.~--}
To study dynamics in finite populations, 
we 
study microscopic stochastic
processes for the dynamics, which lead to
macroscopic equations of motion 
for large populations \cite{helbing,TCH05,TCH06}.
As the microscopic dynamics may
depend on the system, 
the respective biological or behavioral
setup 
may require different interaction
and competition processes.
To demonstrate the robustness of our results,
we consider
different birth-death processes:
the frequency-dependent Moran process 
(MO)
\cite{moran,nowak04,taylor04}, 
and local two-particle interaction 
processes \cite{TCH05,TCH06,BLUME1993!,szabohauert02,TNP06}.

In finite populations with $i$ $R$-players, $j$ $P$-players, and $N-i-j$ $S$-players, the payoffs can be calculated from 
the equations for infinite $N$
by replacing 
$x \to i/(N-1)$,
$y \to j/(N-1)$, and
$z \to (N-i-j)/(N-1)$.
By dividing by $N-1$ and setting the payoff for ties zero, 
we formally exclude self-interactions. 
For the average payoff, we have 
$\langle\pi
\rangle = 
\frac{i}{N} \pi^R
+
\frac{j}{N} \pi^P
+
\frac{N-i-j}{N} \pi^S$.

In the frequency-dependent Moran process,
an individual reproduces
proportional to its fitness. 
Then, the offspring replaces a randomly selected individual  
\footnote{Reproduction proportional to fitness
 implies that $1-w+w \pi$ must be positive. In our case, 
 this leads to $w < 1/2$.}. 
The transition probabilities of the 
possible six hopping events are given by
$(T^{RS}:=T^{R\rightarrow S})$
\begin{eqnarray}
T^{RS}
&=&
\frac{1}{2}
\frac{1-w+w\pi^S }{1-w+w\langle\pi \rangle}
\frac{i}{N} \frac{N-i-j}{N}
\\
T^{SR}
&=&
\frac{1}{2}
\frac{1-w+w\pi^R }{1-w+w\langle\pi \rangle}
\frac{i}{N} \frac{N-i-j}{N}
\end{eqnarray}
and $T^{SP},T^{PS},T^{PR},T^{RP}$ are obtained by cyclic
permutation of $(R,P,S)$ and $(i,j,N-i-j)$.
Fitness is a convex combination of
a constant background fitness (set to 1) and the payoff.
The parameter $w>0$ controls the intensity of selection;
random genetic drift is obtained for $w \to 0$.
For better comparison with the processes below, we
have introduced an additional factor $1/2$.

Selecting an individual proportional to fitness
requires knowledge about every payoff in the population. 
In many cases, it is more realistic to assume that
competition occurs locally between two individuals.
One process of this type is the local update (LU) process
\cite{TCH05,TCH06}, where one individual $b$ is selected randomly
for reproduction, 
compares with annother randomly chosen individual $a$,
and changes strategy with probability
$\frac{1}{2}(1+w(\pi_a -\pi_b))$
\footnote{In Ref.\ \cite{TCH05},
$w$ is replaced by $w/\Delta \pi_{\rm max}$.}.
In general, the reproductive fitness can depend in a nonlinear
way on the payoff difference between two competing agents.
as in
the Fermi process (FP) \cite{BLUME1993!,szabohauert02,TNP06}.

We now
unify the processes
by means of a reproductive function,
\\[-8mm]
\begin{eqnarray}
\Phi_{\rm MO}(b \to a)
&=& 
\frac{1}{2} \, \frac{1-w+w\pi_a }{1-w+w\langle\pi\rangle}\\
\Phi_{\rm LM}(b \to a)
\label{LM}
&=&
(1+w(\pi_a-\langle\pi\rangle))
/2
\\
\Phi_{\rm LU}(b \to a)
&=&
(1+w(\pi_a -\pi_b))
/2
\\
\Phi_{\rm FP}(b \to a)
&=& 
[1+\exp(-w (\pi_a-\pi_b)))]^{-1}.
\end{eqnarray}
\mbox{}\\[-8mm] 
By (\ref{LM})
we
introduce a linearized Moran (LM) process
 as 
first order approximation
of the MO
in the limit
of weak selection, $w\to 0$.
For $w=0$, selection is neutral and the four processes are identical with random drift. 
Note that for the Moran processes, $\Phi(b \to a)$ depends on
the average payoff, whereas for the other processes
two individual payoffs are involved. 
The 
transition
probabilities become
\\[-6mm]
\begin{equation}
T^{ba}=\Phi(b \to a) 
N_a \, N_b  \, /  \,N^2.
\end{equation}
\mbox{}\\[-8mm] 
For all processes, $\Phi(b \to a)$ considers a two-particle (birth-death) process
where an individual with fitness $\pi_a$ compares 
with 
the average fitness
$\langle\pi\rangle$ 
or
with annother individual $\pi_b$.

{\sl Average drift.~--}
For the replicator equation of the symmetric
RPS dynamics,
Eq.~(\ref{constantofmotion}) defines a constant of motion.
As we are interested in the finite-size corrections,
we can use $H$ as an observable for the distance 
to the interior fixed point.
For the processes defined above, the transition
probabilities allow to calculate 
the average change of the constant of motion
within the simplex ($1 \leq i,j \leq N-1$)
as \footnote{Since we are only interested in large N,
a prefactor $(N-1)(N-2)/N^2$ is omitted.
}
\begin{widetext}
\begin{eqnarray}
\label{eq11}
\langle \Delta H  \rangle 
\!\!
&=&
\!\!
\!
\frac{2}{N^5}
\!\sum_{i=1}^{N-1}
\sum_{j=1}^{N-i-1}
\!
\Big[
  i j (N-i-j)
(T^{RS} + T^{SR} + T^{SP} + T^{PS} + T^{PR} + T^{RP})
 \\
&&\!\!
\hphantom{\sum_{i=1}^{N-1} \sum_{j=1}^{N-i-1}\big[}
\hspace*{-3em}
\!
\!
\!
- (i-1)j(N-i-j+1) T^{RS} 
- (i+1)j(N-i-j-1) T^{SR} 
- i(j+1)(N-i-j-1) T^{SP} 
\nonumber\\
&&
\!\!
\hphantom{\sum_{i=1}^{N-1} \sum_{j=1}^{N-i-1}\big[}
\hspace*{-3em}
\!
\!
\!
- i(j-1)(N-i-j+1) T^{PS}  
- (i+1)(j-1)(N-i-j) T^{PR} 
- (i-1)(j+1)(N-i-j) T^{RP} 
\Big].
\nonumber  
\end{eqnarray}
\normalsize
%
\noindent
These expressions
give the exact average drift. 
For the two linear processes, we can approximate the
average drift by replacing the sums by integrals.
Using $x=i/N$, $y=j/N$, 
and $z=1-x-y$, we find in the continuum limit
\begin{eqnarray}
\langle \Delta H \rangle &=&
-\frac{2}{N} \int_{0}^{1} {\rm d}x
\int_{0}^{1-x} {\rm d}y 
\Big[
y(x-z) \left(T^{RS}-T^{SR} \right) 
+x(z-y) \left(T^{SP}-T^{PS} \right)
+z(y-x) \left(T^{PR}-T^{RP} \right)  \Big]
\nonumber
\\   
&&
+\frac{2}{N^2}\int_{0}^{1} {\rm d}x
\int_{0}^{1-x} {\rm d}y 
\Big[
y \left(T^{RS}+T^{SR} \right)
+x \left(T^{SP}+T^{PS} \right)
+z \left(T^{PR}+T^{RP} \right)\Big].
\label{conteq}
\end{eqnarray}
\end{widetext}
From this expression, we can 
perform a comparison of $\langle \Delta H \rangle$ for the different processes.
The neutral case as well as the linear 
cases are analyzed below analytically,
in Fig.\ 1 all processes are compared numerically.

\begin{figure}[htbp]
\noindent
\epsfig{file=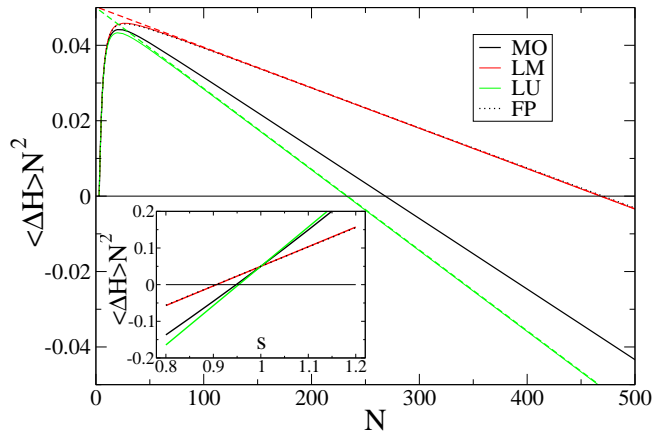,width=0.99\columnwidth} 
\caption{(Color online.)
Drift reversal for the four processes;
the Linearized Moran and  Fermi  process coincide 
within line thickness.
The main panel shows the 
scaling with population size $N$ for
fixed $w=0.45$ and $s=0.8$.
Full lines and the dotted line show the numerical solutions of Eq.\ \eqref{eq11}.
The dashed lines are the analytical
expansions Eqs.\
(\ref{localdrift}) and (\ref{morandrift}).
For small $N$, the continuum approximation to the integral
leads to deviations.
 Inset: 
Average drift for varying payoff matrix with fixed $w=0.45$ and
$N=1000$. The processes intersect in $\Delta H = 1/(20 N^2)$
at $s=1$ ($d=0$) . 
For $N\to\infty$, this reduces to the expectation based on 
the replicator equation, $\Delta H =0$ in $d=0$.
The intersection with the
horizontal line indicates the drift reversal in finite populations.
\label{figreversal}
}
\end{figure}

For neutral evolution, $w=0$, 
all terms of type $T^{RS}-T^{SR}$ vanish, and
we have 
$T^{RS}=T^{SR}=x z/2,$ 
$T^{SP}=T^{PS}=y z/2,$
$T^{PR}=T^{RP}=x y/2,$
i.e.\ for $w=0$, 
\begin{eqnarray}
\langle \Delta H \rangle &=&
\frac{2}{N^2}
\int_{0}^{1} {\rm d}x \int_{0}^{1-x} {\rm d}y  
\frac{6 xyz}{2}
=\frac{1}{20 N^2}. 
\end{eqnarray}

Now we consider the 
linear terms in $w$.
For the linear local update, we find
$T^{RS}-T^{SR}
=w\,xz(\pi^S -\pi^R)$, thus
\begin{eqnarray}
T^{RS}-T^{SR}
 &=& w\, x z( (y-z)+s(y-x)).
\end{eqnarray}
Due to 
$\Phi_{LU}(b \to a)+\Phi_{LU}(a \to b)=1$,
the terms of order $N^{-2}$ in Eq.~(\ref{conteq}) cancel. 
In first and second order 
of $w$, we have 
\begin{eqnarray}
\!\langle \Delta H \rangle_{\rm LU}
= \frac{1}{20N^2}
-\frac{1-s}{420 N}
w.
\label{localdrift}
\end{eqnarray}

For the linearized Moran process, the contributions 
of the average payoff vanish for the drift term, and the
calculation above reproduces with an additional
factor $1/2$, i.e.\
$T^{RS}-T^{SR}=\frac{w}{2}x z(\pi^S - \pi^R)$.
For the diffusion term, we find
\begin{eqnarray}
T^{RS}+T^{SR} &=&xz(1+\frac{w}{2}(\pi^S +\pi^R -2\langle\pi\rangle))
\end{eqnarray}
and
cyclic permutations.
Thus, we get an additional 
contribution to the
diffusion.
In summary, we have
\begin{eqnarray}
\!\langle \Delta H \rangle_{\rm LM}
\label{morandrift}
= \frac{1}{20N^2}
-\frac{1-s}{420N}\left(
\frac{1}{2}
-\frac{1}{N}
\right) w.
\end{eqnarray}

\begin{figure}[htbp]
\noindent
\vspace*{-3mm}
\epsfig{file=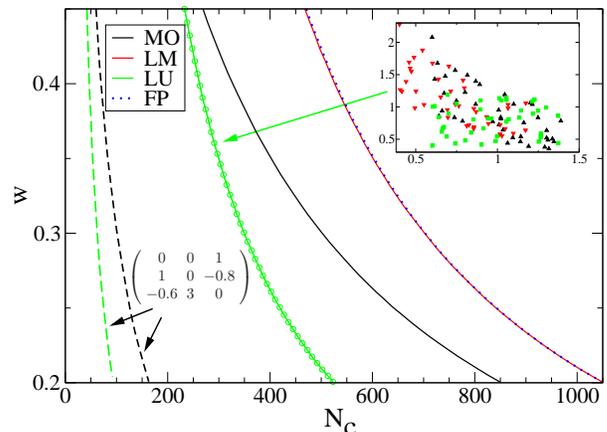,width=0.91\columnwidth} 
\vspace*{-1mm}
\caption{(Color online.)
Critical population size: 
Intensity of selection $w$
versus critical population size $N_c$,
where $\Delta H$ changes its sign
for fixed $s=0.8$.
For $N>N_c$ 
the drift is towards the internal fixed point.
The dashed lines 
refer to the Lizard's payoff matrix 
\cite{sinervo96}
for MO and LU.
Under strong selection $w = 1$ the LU 
reverses at $N_c=20$.
For the restricted random matrix ensemble 
(green circles, see text) under LU
$N_c$ coincides with the 
case of
eq.\ \ref{Mmatrix}.
Each symbol corresponds to a single random payoff matrix.
(inset: scatter plot of payoff entries in
$\blacktriangle$
1$^{\rm st}$,
\Red{$\blacktriangledown$} 2$^{\rm nd}$,
{\Green{{\scriptsize $\blacksquare$}}
3$^{\rm rd}$ column).}
\label{critpopu}}\end{figure}

{\sl 
Biodiversity threshold.~--}
Both Eqs.~(\ref{localdrift}) and (\ref{morandrift})
can change their sign for $s<1$ depending on $N$. 
For all four processes, 
Fig.~\ref{figreversal}
shows the average drift
for different payoffs and 
population sizes.
All processes intersect in 
$+1/(20N^2)$
for $s=1$. In this case, we have a zero-sum game and the average drift is equal to
neutral selection. 
The finite-size correction to the vanishing drift term of neutral selection
arises
 from the difference between our microscopic processes
and symplectic integrators \cite{hofbauer96,sato03}.
In the $N\to\infty$ limit 
we recover that the stability of the
interior fixed point is 
governed by the sign of the
determinant of the payoff matrix,
$d=1-s^3$.
For $s<1$ and $N \to \infty$, 
trajectories spiral inwards.
In finite populations, 
for $s<1$ 
the stochastic motion 
can be reversed
as shown in Fig.~\ref{figreversal}.
The critical population size of the biodiversity threshold is 
$N_c = \frac{21}{w(1-s)}$ 
for the LU and 
$N_c=2+\frac{42}{w(1-s)}$ 
for the LM. 
The biodiversity threshold $N_c$ for the other process is computed numerically in 
Fig.~\ref{critpopu}.
For $N > N_c$,
the $N\to\infty$ behaviour is recovered,
but for $N < N_c$ 
the interior fixed point becomes unstable
and the average drift 
goes towards the boundaries.
So far, we have worked with the specific payoff matrix
Eq.~\eqref{Mmatrix}. Next, we show that this
is a more general phenomenon.
The drift reversal is preserved 
for general cyclic payoffs
with $N_c$ depending on the payoff matrix
\footnote{
Using $H:=-P^{3}$ 
with $P=x^{p_x} y^{p_y} z^{p_z}$ (where $p_x$, $p_y$ and $p_z$ are the coordinates of the fixed point)
and normalization constant  $\Sigma$, see 
 \cite{hofbauer98} for details.}.
As an example, we consider the payoff matrix of the Lizard system \cite{sinervo96}. 
In this case, we find $N_c \geq 20$, see Fig.\ \ref{critpopu}.
More general, we can also address random payoff matrices with cyclic dominance. 
Confining a random matrix ensemble
(Fig.\ \ref{critpopu}) onto a suitable
2-dim submanifold 
\footnote{With two 
payoff matrix entries
equiprobable within a
finite interval; 
remaining nondiagonal entries
were computed such that 
the coordinates of the fixed point and $\Sigma$ 
keep fixed values.}, 
we show that 
 $N_c(w)$ 
depends only on the location
of the fixed point
${\boldsymbol p}$, its normalization
constant $\Sigma$
and 
the determinant $d$,
i.e.\ on 4 parameters.
Hence the phenomenon is generic for
arbitrary cyclic payoff matrices with $d>0$.

{\sl To conclude},
cyclic coevolution 
in biological or social dynamics
highlights the importance to study
finite population effects 
and the underlying microscopic dynamics.
Recently the influence of the finiteness of the
population has been widely discussed for 2$\times$2 
games
\cite{nowak04,taylor04,TCH05,TCH06,CT05}.
But also in cyclic dynamics
the finiteness of the population 
can
modify the stability conditions 
derived from the replicator equation.
In this Letter, we have shown that biodiversity threshold of
the population size
occurs in generic cyclic 3$\times$3 games 
for a positive determinant of the payoff matrix. 
Such a positive determinant has been found 
in the Lizard system
\cite{sinervo96,Zam00}.
Thus, stability of the coexistence fixed point
can be obtained for sufficiently large 
populations, preserving biodiversity
even in non-spatial systems 
or under strong mobility.
In contrast, experiments in
well-mixed populations of the E.coli system 
indicate a negative determinant of the payoff matrix.

 Here we have demonstrated how nonzero-sum payoffs
can change
 the stability even in well-mixed systems.
Thus in biological ---~and corresponding
social or economic~--- systems, 
payoff matrix, population structure, population size
and the microscopic update mechanism
determine the fate of extinction or coexistence.

\end{document}